\documentclass[amsmath,amssymb,aps,pre,floatfix,twocolumn]{revtex4-1}

\usepackage{graphicx}
\usepackage{dcolumn}
\usepackage{bm}
\usepackage{subfigure}  
\newcommand{\LL}{\textsf{L}}

\usepackage{exscale}

\usepackage{xcolor}
\usepackage[plainpages=false,pdfpagelabels,colorlinks=true,linkcolor=red,urlcolor=blue,citecolor=blue,pdftitle={},pdfauthor={},pdfdisplaydoctitle=true,pdfduplex=DuplexFlipLongEdge]{hyperref}

\definecolor{darkred}{rgb}{0.90,0.2,0.2}

\begin{document}

\preprint{APS/123-QED}

\title{{\LL}-based numerical linked cluster expansion for square lattice models}
\author{Mahmoud Abdelshafy}

\affiliation{Department of Physics, The Pennsylvania State University, University Park, Pennsylvania 16802, USA}
\author{Marcos Rigol}
\affiliation{Department of Physics, The Pennsylvania State University, University Park, Pennsylvania 16802, USA}

\date{\today}

\begin{abstract}
We introduce a numerical linked cluster expansion for square-lattice models whose building block is an {\LL}-shape cluster. For the spin-1/2 models studied in this work, we find that this expansion exhibits a similar or better convergence of the bare sums than that of the (larger) square-shaped clusters, and can be used with resummation techniques (like the site- and bond-based expansions) to obtain results at even lower temperatures. We compare the performance of weak- and strong-embedding versions of this expansion in various spin-1/2 models, and show that the strong-embedding version is preferable because of its convergence properties and lower computational cost. Finally, we show that the expansion based on the {\LL}-shape cluster can be naturally used to study properties of lattice models that smoothly connect the square and triangular lattice geometries. 
\end{abstract}

\maketitle

\section{Introduction}
Strongly-correlated quantum lattice models play an important role in our understanding of unusual properties of materials, such as insulating behaviors in regimes that would be metallic in the absence of interactions, magnetism, and (possibly) high-temperature superconductivity~\cite{dagotto_94, georges_96, imada_98, Jaklic_2000, lee_06}. They are challenging to study analytically because of the absence of small parameters to carry out perturbative expansions, and because strong correlations and quantum fluctuations preclude the development of reliable mean-field field theory descriptions. They are also challenging to study computationally using full exact diagonalization calculations because the size of the Hilbert space grows exponentially with the number of lattice sites. A wide range of computational approaches has been developed to overcome the latter challenge, such as Quantum Monte Carlo (QMC) techniques~\cite{scalapino_81, sandvik_10, Pollet_2012, Li_QMC_2019}, whose applicability is limited by the sign problem \cite{loh_90, henelius_00, troyer_05}; the density matrix renormalization group (DMRG) technique~\cite{white_92, white_93, schollwock_05}, most efficient for spin chains; and series expansions methods~\cite{domb_green_book_74, oitmaa_hamer_book_06}, just to name a few.

Series expansion methods generally contain a small parameter, e.g., for the commonly used high-temperature expansions the small parameter is the inverse temperature~\cite{domb_green_book_74, oitmaa_hamer_book_06}. Consequently, high-temperature expansions fail to converge at low temperatures. This failure is independent of the nature of the low-temperature state, which, e.g., could exhibit short-range correlations or a slow build-up of correlations as the temperature is lowered. In order to overcome this limitation of high-temperature expansions, as well as potentially similar limitations of other expansions involving small parameters of the Hamiltonian, in Refs.~\cite{rigol_bryant_06, rigol_bryant_07a, rigol_bryant_07b} it was shown that one can use numerical linked cluster expansions (NLCEs). NLCEs allow one to compute the expectation values of extensive observables per site in the thermodynamic limit adding contributions from increasingly large connected clusters. No small parameter is assumed a priori when carrying out NLCEs, but they only converge when the correlations in the system are smaller or of the same order as the sizes of the clusters considered. At a fixed temperature when the cluster sizes exceed the correlation length, NLCE results have been shown to converge to the thermodynamic limit ones exponentially fast as the size of the clusters considered is increased~\cite{iyer_15}. 

NLCEs have been used to study a wide range of lattice models in thermal equilibrium to understand properties of materials~\cite{rigol_singh_07a, rigol_singh_07b, khatami_singh_11, singh_12, khatami_helton_12, applegat_12, hayre_13, sherman_16, sherman_18, benton_18a, benton_18b, schafer_20, sarkis_20, singh_22} and of ultracold gases in optical lattices~\cite{khatami_rigol_11, russel_15, duarte_15, cheuk_16, brown_17, nichols_19}. They have also been used to study entanglement~\cite{kallin_13, stoudenmire_14, devakul_15, sherman_devakul_16, hayward_17, pardini_19}, thermodynamic properties of disordered systems~\cite{tang_iyer_15b, mulanix_19}, observables after equilibration following quantum quenches~\cite{rigol_14a, wouters_14, rigol_14b, tang_iyer_15a, rigol_16, piroli_17, mallayya_17}, and quantum dynamics~\cite{mallayya_18, white_17, dekavul_18, mallaya_rigol_19, mallaya_19, gan_20, richter_20}. Two important characteristics of NLCEs that have been explored and successfully used in the various applications mentioned above are the freedom to carry out the expansions using different building blocks, as well as the fact that one can use resummation techniques to extend the convergence of NLCEs beyond that provided by the bare sums, namely, to regimes in which the extent of the correlations exceeds the largest clusters considered.

In this work, we introduce a numerical linked cluster expansion for square-lattice models whose building block is an {\LL}-shape cluster. The number of clusters (the number of sites in each cluster) at each order of the {\LL}-based NLCE increases more slowly (more rapidly) than for the site- and bond-based expansions. Consequently, in the ``{\LL} expansion'' one can include clusters with more sites than those that can be included in the ``site'' and ``bond'' expansions. On the other hand, the number of clusters (the number of sites in each cluster) at each order of the {\LL} expansion increases more rapidly (more slowly) than for the square-based expansion. Consequently, in the {\LL} expansion one can include more clusters than those that can be included in the ``square expansion''. The site, bond, and square expansions for the square lattice were introduced, and used to study thermodynamic properties of spin-1/2 models in thermal equilibrium, in Ref.~\cite{rigol_bryant_07a}. Remarkably, for the spin-1/2 models considered here, the {\LL} expansion exhibits a convergence of the bare sums that is either similar or faster than that of the square expansion. They both converge at lower temperatures than the site and bond expansions. In addition, for the {\LL} expansion one can use resummation techniques to extend the convergence beyond that of the bare sums. This is something that, so far, has been consistently achieved only in the context of the site- and bond-based expansions. We compare the performance of weak- and strong-embedding versions of the {\LL} expansion, and show that the strong-embedding version is the better choice because of its convergence properties and lower computational cost. Finally, we show that the {\LL} expansion can be used to study properties of lattice models that smoothly connect the square and triangular lattice geometries. 

The presentation is organized as follows. In Sec.~\ref{sec:Ham}, we introduce the spin-1/2 model Hamiltonians studied in this work. A short summary of NLCEs, the resummation techniques used, as well as the observables that we study and how we gauge the convergence of the NLCE results, is provided in Sec.~\ref{sec:methods}. The {\LL} expansion is introduced in Sec.~\ref{sec:lnlce}. The numerical results obtained for spin-1/2 Ising, XX, and Heisenberg models are reported in Secs.~\ref{sec:ising},~\ref{sec:xx}, and~\ref{sec:heisenberg}, respectively. A summary of our results is presented in Sec.~\ref{sec:summary}. 

\section{Model Hamiltonians}\label{sec:Ham}

\begin{figure}[!b]
    \centering
    \includegraphics[width=0.9\columnwidth]{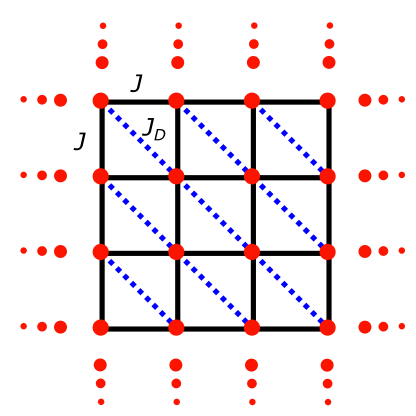}
    \caption{Lattice geometry studied in this work. The diagonal bonds allow us to study models that smoothly connect the square and triangular lattices.}
    \label{fig:sq_lattice}
\end{figure}

\begin{figure}[!t]
    \centering
    \includegraphics[width=0.6\columnwidth]{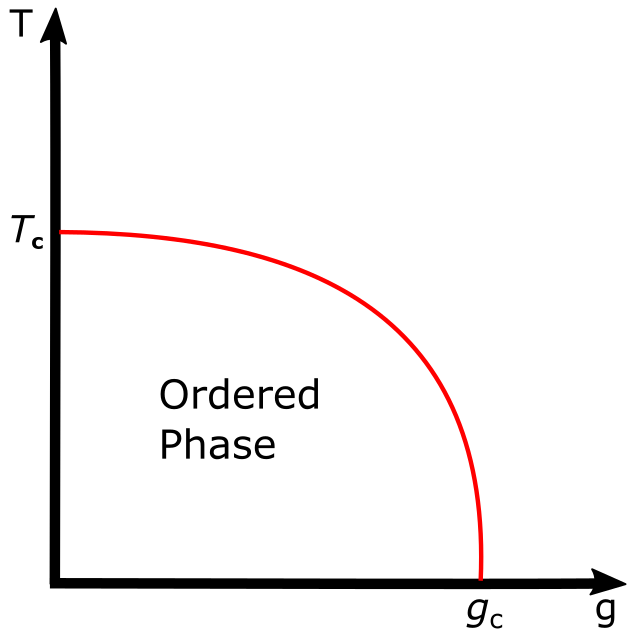}
    \caption{Sketch of the phase diagram, on the $T-g$ plane, for the transverse-field Ising model in the square lattice. In the phase diagram, $T_c\approx 0.57$ and $g_c\approx 1.52$.}
    \label{fig:TFIM_phase}
\end{figure}

We consider three translationally invariant spin-1/2 Hamiltonians in an infinite square lattice (see Fig.~\ref{fig:sq_lattice}). The first one is the Ising model in a transverse field
\begin{equation}
\hat{H}=J\sum_{\langle {\bf i},{\bf j}\rangle} \hat{S}^{z}_{\bf i}\hat{S}^{z}_{\bf j} + g\sum_{\bf i} \hat{S}^{x}_{\bf i} \, ,
\label{eq:HIsing}
\end{equation}
where $\hat{S}^{z}_{\bf i}$ ($\hat{S}^{x}_{\bf i}$) are the $z$ ($x$) spin-$\frac{1}{2}$ operators at site ${\bf i}$, $\langle {\bf i},{\bf j}\rangle$ denotes pairs of nearest neighbors sites in the lattice, and we set $J=1$ as our energy scale. For $g = 0$, the Hamiltonian in Eq.~\eqref{eq:HIsing} is that of the classical Ising model, which is exactly solvable~\cite{Onsager_1944} and whose solution we will use to quantify the convergence of the {\LL} expansion. The classical Ising model develops long-range order below a critical temperature $T_c\approx 0.57$. For $g \neq 0$, the Hamiltonian in Eq.~\eqref{eq:HIsing} is that of the {\it quantum} transverse-field Ising model, which is not exactly solvable but has been studied in detail using numerical simulations. At zero temperature, this model exhibits two extended phases, an ordered phase for $g< g_c\approx 1.52$ and a disordered phase for fields $g> g_c$~\cite{du_Croo_TFIM_1998, Youjin_TFIM_2002}. For fields below the critical field $g_c$, the transverse field Ising model develops long-range order below a critical temperature that depends on the value of $g$~\cite{TFIM_phase_diagram}. A sketch of the phase diagram for this model is shown in Fig.~\ref{fig:TFIM_phase}.

The other two models that we study are the XX (also referred to sometimes as XY) and Heisenberg models
\begin{equation}
\hat{H}=J\sum_{\langle {\bf i},{\bf j}\rangle} \left( \hat{S}^x_{\bf i} \hat{S}^x_{\bf j} + \hat{S}^y_{\bf i} \hat{S}^y_{\bf j}  + \Delta \hat{S}^{z}_{\bf i}\hat{S}^{z}_{\bf j} \right),
\label{eq:HHeis}
\end{equation}
where, $\Delta=0$ corresponds to the XX model and $\Delta=1$ corresponds to the Heisenberg model. As for the Ising case, we set $J=1$ to be our energy scale. The XX model has $U(1)$ symmetry and the Heisenberg model has $SU(2)$ symmetry. As a result, because of the Mermin–Wagner theorem, these models only develop long-range order at zero temperature. The XX model exhibits a Berezinskii-Kosterlitz-Thouless (BKT) transition~\cite{Berezinsky:1972rfj, Kosterlitz_1973} at $T_{KT} \approx 0.343$~\cite{harada_kawashima_97, XY_Tkt, carrasquilla_rigol_12}, below which the system exhibits power-law decaying correlations. The XX and Heisenberg models in the square lattice are not exactly solvable, but they can be efficiently simulated using QMC techniques because they are not frustrated and, hence, do not suffer from the sign problem. Consequently, their properties can be computed with high accuracy.

In addition to studying the spin-1/2 transverse-field Ising, XX, and Heisenberg models in the square lattice geometry, we study the thermodynamic properties of those models as they transition between the square and triangular lattice geometries. This can be done by adding bonds along one of the diagonals of the square lattice, as shown in Fig.~\ref{fig:sq_lattice}. We label the strength of the interactions along those bonds as $J_D$. For $J_D=0$ we have the square lattice geometry, while for $J_D=J$ we have the triangular lattice geometry. For $J_D \rightarrow \infty$, our model Hamiltonians become those of disconnected chains, so studying $J_D>J$ (which we do not do here) allows one to explore the dimensional crossover between the triangular lattice and disconnected spin-1/2 chains.

\section{A short summary of NLCEs}\label{sec:methods}

In NLCEs (see Ref.~\cite{Tang_2013} for a pedagogical introduction) extensive observables per lattice site $\mathcal{O}/N$, in an arbitrary lattice with $N$ sites, can be computed as the sum over the contributions from the connected clusters that can be embedded on the lattice:
\begin{equation}\label{eq:nlce}
\mathcal{O}/N=\sum_{c} L(c)\times W_{\mathcal{O}}(c),
\end{equation}
where $L(c)$ is the embedding factor, which counts the number of ways per site that cluster $c$ can be embedded on the lattice, and $W_{\mathcal{O}}(c)$ denotes the weight of observable $\mathcal{O}$ in cluster $c$. The weights in Eq.~\eqref{eq:nlce} are calculated using the inclusion-exclusion principle:
\begin{equation}\label{eq:weight}
W_{\mathcal{O}}(c)=\mathcal{O}(c)- \sum_{s\subset c} W_{\mathcal{O}}(s),
\end{equation}
with $W_{\mathcal{O}}(c)=\mathcal{O}(c)$ for the smallest cluster. This recursive relation ensures that the weight of cluster $c$ includes only the contribution to the observable arising from correlations between all the sites in the particular geometry of cluster $c$. The expectation value $\mathcal{O}(c)$ of an observable for cluster $c$, with a many-body density-matrix operator $\hat\rho_c$, is calculated using full exact diagonalization: 
\begin{equation}
\mathcal{O}(c)=\text{Tr}\left(\hat{O}\,\hat\rho_c\right).
\end{equation}
The latter is the bottleneck of the NLCE computations, because the sizes of the matrices involved in the full exact diagonalization calculations --- which we carry out using LAPACK --- scale exponentially with the sizes of the clusters considered. Identifying all clusters that give the same expectation value $\mathcal{O}(c)$, which we call topologically equivalent clusters, significantly reduces the computational cost of evaluating the sum in Eq.~\eqref{eq:nlce}. 

The convergence of the series in Eq.~(\ref{eq:nlce}) is automatically guaranteed for lattices with a finite number of sites $N$. When all subclusters are included, using the relation in Eq.~\eqref{eq:weight} and reordering the terms, one gets $\mathcal{O}(c)$~\cite{Tang_2013}. However, to obtain the thermodynamic limit ($N \rightarrow \infty$) result, which is the limit of interest to describe the properties of materials, the series must be truncated after including a finite number of terms. The truncated sum is expected to provide an accurate prediction for the thermodynamic limit result when the extent of the connected correlations involved for the given observable is smaller than, or of the same order as, the sizes of the clusters included in the sum. In that regime, one expects the weights $W_{\mathcal{O}}(c)$ to decrease rapidly with increasing the size of the clusters. In fact, an exponential convergence has been observed in unordered phases in thermal equilibrium~\cite{iyer_15}. 

\subsection{Building blocks}\label{sec:blocks}

A striking flexibility of NLCEs is that, for any given lattice geometry, one can use different building blocks to carry out different expansions. For the square lattice, in Ref.~\cite{rigol_bryant_07a} it was shown that one can use bonds, sites, and corner-sharing squares as building blocks for NLCEs to study thermal equilibrium properties of spin models. More recently, a rectangle expansion was used to study quantum dynamics~\cite{gan_20, richter_20}. The number of clusters in the bond and site expansions grows very rapidly with the number of bonds and sites, respectively, in the clusters. As a result, because of the large number of clusters, the series for spin-1/2 models needs to be truncated at cluster sizes that are smaller than those that can be fully diagonalized in current computers. On the other hand, the square/rectangle expansions exhibit a number of clusters that grows slowly with the number of sites in the clusters. This results in a number of clusters that is not too large for the cluster sizes that can be fully diagonalized for spin-1/2 models in current computers. In this case, the limit in the number of terms in the expansion is set by the largest clusters that one can diagonalize. As we will show, each expansion has its advantages and disadvantages, which are associated to: (i) the convergence of the bare sums, and (ii) the possibility of using resummation techniques.

\subsection{Resummations}\label{sec:resum}

Beyond the regime in which the truncated bare sums converge, one can use resummation techniques to make predictions in regimes in which the extent of the correlations exceeds the cluster sizes. The two most commonly used resummation techniques are Wynn's and Euler algorithms~\cite{rigol_bryant_06, rigol_bryant_07a, rigol_bryant_07b}. 

To implement resummation techniques, one needs to group together the contributions of all clusters that share a particular property, e.g., that have $n$ sites, or $n$ bonds, or $n$ squares,
\begin{equation}\label{S_n}
 S_{n}=\sum_{\{c^n_{\ell}\}} \mathcal{L}(c^{n}_{\ell})\times W_{\mathcal{O}}(c^n_{\ell}).
\end{equation}
Using those partial sums, one can rewrite the truncated sums for Eq.~\eqref{eq:nlce} as
\begin{equation}
	\mathcal{O}_l/N= \sum_{n=1}^{l} S_{n}.
\end{equation}
The resummation algorithms utilize the finite sequence $\{\mathcal{O}_l\}$ to predict the result for $\mathcal{O}_{l\rightarrow\infty}$, i.e., for the observable in the thermodynamic limit. 

In the Wynn's ($\epsilon$) resummation algorithm~\cite{rigol_bryant_06, rigol_bryant_07a, rigol_bryant_07b}, $\epsilon_n^{(k)}$ is defined as
\begin{eqnarray}
&&\epsilon_n^{(k)}=\epsilon_{n+1}^{(k-2)}+\frac{1}{\epsilon_{n+1}^{(k-1)}-\epsilon_n^{(k-1)}},\\
&&\text{with\quad}\epsilon_n^{(-1)}=0, \quad \epsilon_n^{(0)}=\mathcal{O}_n,\nonumber 
\end{eqnarray}
where $k$ denotes the number of Wynn resummation ``cycles''. Only even entries $\epsilon_n^{(2k')}$ (with $k'$ an integer) are expected to converge to the thermodynamic limit result. The new sequence generated after two cycles has two fewer terms, limiting the maximum possible number of cycles. The estimate after $2k'$ cycles is given by
\begin{equation}
\text{Wynn}_{k'}(\mathcal{O}/N)=\epsilon_{l_\text{max}-2k'}^{2k'}.
\end{equation} 
where we call $k'$ the Wynn resummation ``order'', and $l_\text{max}$ is the largest value of $l$ for our sequence $\{\mathcal{O}_l\}$. One can see that the larger the value of $l_\text{max}$ the higher the order at which Wynn's resummation algorithm can be carried out. Hence, the better the chance of extending the NLCE convergence beyond that of the bare sums.
 
Another resummation scheme that can accelerate the convergence of {\it alternating} series, i.e., series in which the partial sums $S_{n}$ alternate signs, is the Euler transformation. With this algorithm, the thermodynamic limit result is obtained using the formula
\begin{eqnarray}
&&\text{Euler}_k(\mathcal{O}/N)=\sum_{n=0}^{l_\text{max}-k}S_{n}+(-1)^{l_\text{max}-k+1}\sum_{n=0}^{k}\frac{1}{2^{n+1}} T^{l_\text{max}}_{k,n},  \nonumber \\
&&\text{where\quad}
T^{l_\text{max}}_{k,n}=\sum_{j=0}^{n}{n\choose j}S_{l_\text{max}-k+n-j+1}.\label{eq:euler}
\end{eqnarray}
To improve the outcome of the Euler transformation, one can vary $k$ as needed.

By comparing the results of different resummation techniques and orders one can gauge whether they are providing meaningful results. 

\subsection{Observables and convergence}\label{sec:observables}

We study systems that are in thermal equilibrium at finite temperature, and which are described by the Hamiltonians in Sec.~\ref{sec:Ham}. For those systems, we carry out calculations in the grand canonical ensemble at zero chemical potential, namely, given $\hat{H}$ and a temperature $T$, the many-body density matrix is taken to be
\begin{equation}\label{eq:rho}
    \hat{\rho} = \frac{1}{Z} \exp \left( -\frac{\hat{H}}{k_BT} \right)\, ,
\end{equation}
where
\begin{equation}
Z=\text{Tr} \left[ \exp \left( -\frac{\hat{H}}{k_BT}\right) \right].
\end{equation}
In what follows, we set the Boltzmann constant $k_B=1$. 

We report results for a set of three extensive observables. The total energy per site:
\begin{equation}
E=\frac{1}{N}\text{Tr}[\hat{H}\hat\rho]\, ,
\end{equation}
the specific heat per site: 
\begin{equation}
C_v=\frac{1}{N}\frac{\text{Tr}(\hat{H}^2\hat\rho)-[\text{Tr}(\hat{H}\hat\rho)]^2}{T^2}\, ,
\end{equation}
and the entropy per site: 
\begin{equation}
S=-\frac{1}{N} \text{Tr}(\hat\rho \ln \hat \rho) = \frac{1}{N} \ln Z + \frac{E}{T}\, .
\end{equation}

To study the convergence of the truncated bare sums for an observable $\mathcal{O}$ in general, we calculate the normalized difference
\begin{equation}\label{rel_error_TH}
\Delta_{l}(\mathcal{O})=\left|\frac{\mathcal{O}_{l_\text{max}}-\mathcal{O}_{l}}{\mathcal{O}_{l_\text{max}}}\right|.
\end{equation}
If the $l^\text{th}$ normalized difference $\Delta_l(\mathcal{O})$ reaches machine precision, we conclude that the NLCE result for the $l^\text{th}$ and higher orders has converged to the thermodynamic limit result (within machine precision). For models for which an exact result for $\mathcal{O}$ is known, we replace $\mathcal{O}_{l_\text{max}}$ in Eq.~\eqref{rel_error_TH} by the exact result.

\section{{\LL} expansion}\label{sec:lnlce}

\begin{figure}[!b]
\centering
\includegraphics[width=0.45\columnwidth]{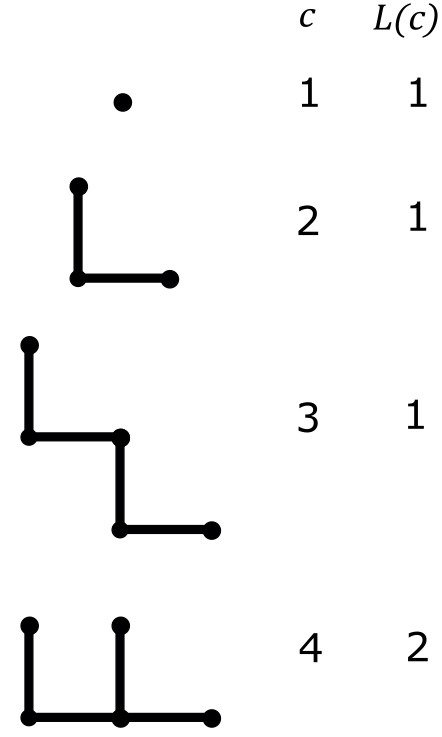}
\caption{Clusters that appear in the first three orders of the {\LL} expansion.}\label{fig:L_clusters}
\end{figure}

In this section we introduce the {\LL} expansion, whose building block is an {\LL}-shape cluster. We find that, for the spin-1/2 models studied here, such an expansion shares the advantages of the square-based NLCE over the site- and bond-based ones, because the bare sums in the {\LL} expansion exhibit a significantly better convergence at lower temperatures than the bare sums of the site and bond expansions (the convergence of the {\LL} expansion is also better in most cases considered than that of the square expansion). The {\LL} expansion also shares the advantages of the site- and bond-based NLCEs over the square-based one, because resummation techniques can be used to significantly extend convergence to temperatures that are lower than those accessible via the bare sums of any of the expansions considered.

The clusters appearing in the first 3 orders of the {\LL} expansion are shown, along with their embedding factors, in Fig.~\ref{fig:L_clusters}. The first, and smallest, cluster of the {\LL} expansion, which is the same as for the other expansions mentioned, is the single site. This cluster determines the infinite-temperature value of various observables, such as the entropy, and constitutes the zeroth order of the {\LL} expansion. The second cluster is the building block of our expansion, an {\LL}-shape cluster. There is only one such cluster per site in the square lattice. This cluster constitutes the first order of the {\LL} expansion. The second order includes two clusters in which the {\LL} building blocks share a site. As for the corner-sharing square expansion previously mentioned, larger clusters are constructed by adding building block clusters that share sites (not bonds) with previously constructed clusters. Note that the two clusters in the third order of the expansion have different values of $L(c)$, $c=3$ can be embedded only one way per site whilst $c=4$ has $L(c=4)=2$ as the two {\LL} blocks can be arranged horizontally or vertically.

\begin{table}[!t]
\caption{Total number of clusters (third column) and the number of topological distinct clusters (second column) in the weak-embedding {\LL} expansion versus the number of {\LL}s in the clusters (first column).}
\label{Table_ngraphs_W} 
\begin{ruledtabular}
\begin{tabular}{r r r}
No.~{\LL}s	 &No.~topological clusters	 & Total number of clusters \\
\hline
0           &1           &1      \\
1           &1           &1      \\
2           &2           &3      \\   
3           &6           &11     \\  
4           &19          &44     \\  
5           &68          &186    \\ 
6           &256         &814    \\ 
7           &1018        &3652   \\ 
8           &4162        &16689  \\ 
9           &17423       &77359  \\
10          &74073       &362671 \\
\end{tabular}
\end{ruledtabular}	
\end{table}

\begin{figure}[!b]
\centering
\includegraphics[width=0.85\columnwidth]{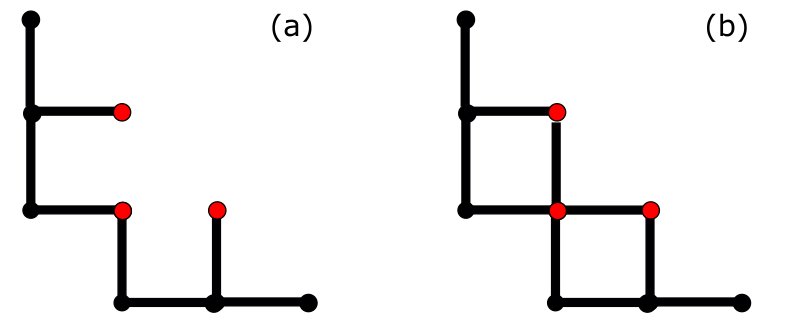}
\caption{(a) A cluster with four {\LL}s that only appears in the weak-embedding {\LL} expansion. (b) A cluster with five {\LL}s that appears both in the weak-embedding and strong-embedding {\LL} expansions.}\label{fig:strong_embeddings}
\end{figure}

The order of the {\LL} expansion can be defined to be the number of {\LL} building blocks starting, as mentioned before, from the zeroth order for the single site. As more {\LL} building blocks are added, one can choose to carry out ``weak-embedding'' and ``strong-embedding'' versions of the {\LL} expansion. Those versions parallel the bond and site expansions, respectively. In the site expansion, all possible bonds are placed between existing sites, e.g., in such an expansion there cannot be an open square cluster (namely, a cluster with 4 sites forming a square and only 3 bonds), which is a valid cluster in the bond expansion~\cite{rigol_bryant_07a}. Similarly, in the strong-embedding {\LL} expansion, all possible {\LL} building blocks must be placed between existing sites. In other words, a cluster with 3 sites making an {\LL} without being linked by bonds can only exist in a weak-embedding version. An example of a cluster with four {\LL}s that is only present in the weak-embedding {\LL} expansion is shown in Fig.~\ref{fig:strong_embeddings}(a). The cluster with five {\LL}s shown in Fig.~\ref{fig:strong_embeddings}(b) then appears both in the weak and strong embedding expansions. Consequently, the number of clusters increases more slowly in the strong-embedding {\LL} expansion and those clusters end up being more ``compact'' than those of the weak-embedding {\LL} expansion.

\begin{table}[!t]
\caption{Total number of clusters (third column) and the number of topological distinct clusters (second column) in the strong-embedding {\LL} expansion versus the number of {\LL}s in the clusters (first column).}
\label{Table_ngraphs_S} 
\begin{ruledtabular}
\begin{tabular}{r r r}
No.~{\LL}'s	 &No.~topological clusters	 & Total number of clusters \\
\hline
0           &1           &1      \\
1           &1           &1      \\
2           &2           &3      \\   
3           &6           &11     \\  
4           &18          &41     \\  
5           &61          &153    \\ 
6           &202         &573    \\ 
7           &700         &2162   \\ 
8           &2429        &8238   \\ 
9           &8608        &31696  \\
10          &30734       &122986 \\
\end{tabular}
\end{ruledtabular}	
\end{table}

The total number of clusters in the weak and strong embedding versions of the {\LL} expansion are shown in the third column of Tables~\ref{Table_ngraphs_W} and~\ref{Table_ngraphs_S}, respectively, as functions of the number of {\LL}s in the clusters -- the first column in those tables. By the total number of clusters we mean the sum of $L(c)$ for all clusters with any given number of {\LL}s. In order to reduce the computational cost of the numerical calculations, we start by finding the clusters that are related by lattice symmetries, e.g., the last cluster in Fig.~\ref{fig:L_clusters}, which gives the same contribution to thermodynamic quantities as the one with the two {\LL}s arranged vertically. Only one needs to be diagonalized. As a second step, we find all clusters that are not related by lattice symmetries but whose Hamiltonians are the same, we refer to those clusters as topologically equivalent clusters and only diagonalize one of them. The second columns in Tables~\ref{Table_ngraphs_W} and~\ref{Table_ngraphs_S} show the number of topologically inequivalent clusters for the weak and strong embedding versions of the {\LL} expansion, respectively.

Comparing the number of topologically distinct clusters for the {\LL} expansion, in Tables~\ref{Table_ngraphs_W} and~\ref{Table_ngraphs_S}, and the number of topologically distinct clusters for the bond, site, and square expansions reported in Ref.~\cite{rigol_bryant_07a} one can see that, at any given order, the {\LL} expansion contains significantly fewer topologically distinct clusters than the bond and site expansions, and significantly more than the square expansion. In addition, the number of orders that can be calculated for the {\LL} expansion is larger than for the square expansion. Hence, resummation techniques are more likely to improve the convergence of the {\LL} expansion than that of the square expansion~\cite{rigol_bryant_07a, khatami_rigol_11_sq}. We show results that indicate that the {\LL} expansion provides a ``sweet spot'' for NLCE calculations in the square lattice. 

\begin{figure}[!t]
\centering
\includegraphics[width=0.85\columnwidth]{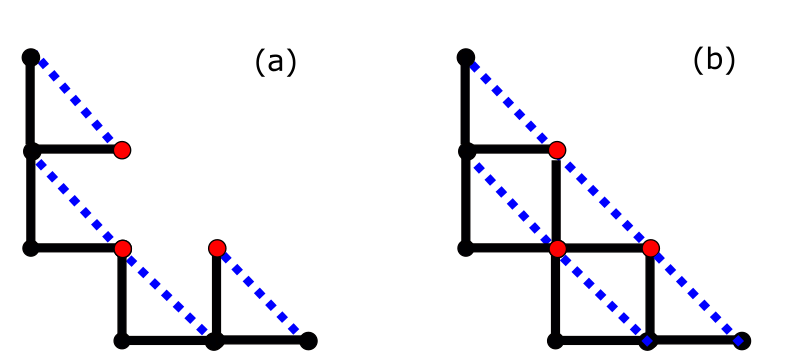}
\caption{The same clusters as in Fig.~\ref{fig:strong_embeddings} after adding a bond (dashed blue) along the diagonal of each {\LL} building block. When all the bonds have an identical strength, the resulting lattice geometry is that of the triangular lattice.}\label{fig:triangular}
\end{figure}

The {\LL} expansion also makes possible something that none of the previously mentioned expansions for the square lattice do. If one adds a bond along the diagonal of each {\LL} building block, by changing the strength $J_D$ of that bond between 0 and $J$, one can study models that smoothly connect the square ($J_D=0$) and triangular ($J_D=J$) lattice geometries (see Fig.~\ref{fig:triangular}). In fact, for $J_D=J$, the resulting weak-embedding {\LL} expansion is nothing but the corner-sharing triangle expansion introduced in Ref.~\cite{rigol_bryant_07a} for the triangular lattice. This can be seen by comparing the numbers of clusters in Table~\ref{Table_ngraphs_W} to those reported in Ref.~\cite{rigol_bryant_07a} for the triangle expansion (after correcting for an incorrect factor of 3 in the table in the latter reference). The strong-embedding {\LL} expansion is then a new expansion that, for $J_D=J$, can be used to study models in the triangular lattice.

\section{Ising models}\label{sec:ising}

In this section, we use the {\LL} expansion for Ising models to explore how our observables of interest behave in the square lattice as the temperature is changed. We also study how some of those observables behave as one transitions between the square and the triangular lattice by changing the strength $J_D$ of the diagonal bonds, and as one transitions between the classical and quantum regimes by making $g\neq 0$. When available, we compare our numerical results to exact analytical results. All the calculations are done in the grand canonical ensemble at zero chemical potential, using the density operator written in Eq.~\eqref{eq:rho}.

\subsection{Ising model}\label{sec:IS}

\begin{figure}[!b]
\includegraphics[width=.98\columnwidth]{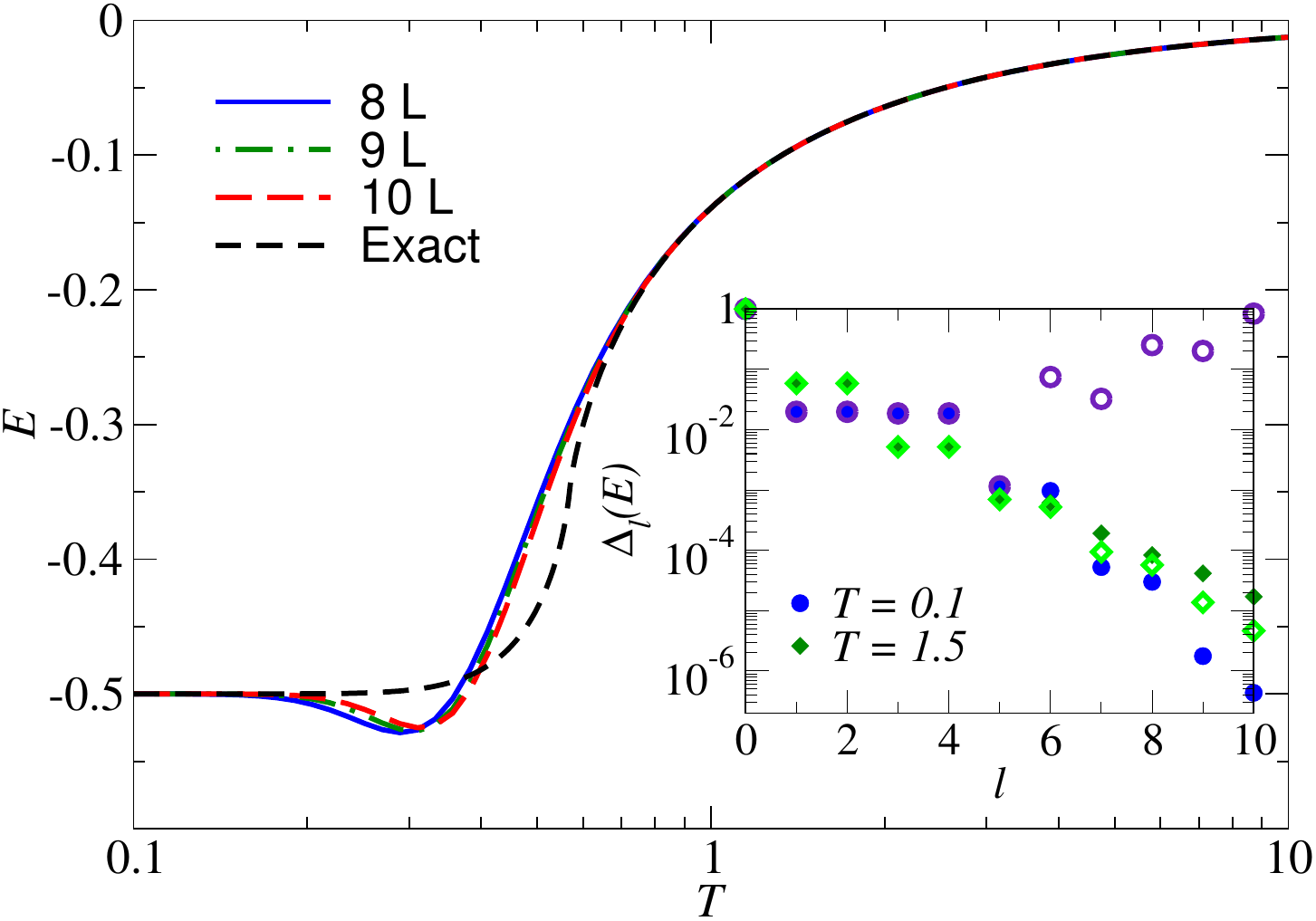}
\caption{Energy per site vs $T$ for the strong-embedding {\LL} expansion, with up to 8, 9, and 10 {\LL}s, and the exact energy for the square lattice Ising model. (inset) Normalized difference between the energy obtained using the strong-embedding (filled symbols) and weak-embedding (empty symbols) {\LL} expansions and the exact result plotted as functions of the order of the {\LL} expansion. We show results for two temperatures, one above and one below $T_c$.}\label{fig:IS_sq_E}
\end{figure}

On a square lattice, the Ising model [$g = 0$ in Eq.~\eqref{eq:HIsing}] was solved by Onsager in 1944~\cite{Onsager_1944}. As mentioned in Sec.~\ref{sec:Ham}, this model exhibits a phase transition between a disordered phase at $T>T_c$ and an ordered phase at $T<Tc$. In Fig.~\ref{fig:IS_sq_E}, we show results of the strong-embedding {\LL} expansion for the energy vs temperature, obtained using clusters with up to 8, 9, and 10 {\LL}s, as well as the exact analytical result. The NLCE results are indistinguishable from the exact ones, signaling convergence, at temperatures $T\lesssim 0.2$ and $T\gtrsim 0.8$. At those temperatures, both below and above $T_c$, we find that the accuracy of the strong-embedding {\LL} expansion increases exponentially with the order of the expansion. This is shown, at two temperatures (one above and one below $T_c$), by the filled symbols in the inset in Fig.~\ref{fig:IS_sq_E}. The main panel in Fig.~\ref{fig:IS_sq_E} also shows that, with increasing the order of the strong-embedding {\LL} expansion, the temperatures over which the NLCE and the exact results agree, i.e., over which the NLCE converges, approach $T_c$ for temperatures both below and above $T_c$. For a model with an unknown finite-temperature phase transition, such a behavior could be used as a way to identify the critical region in which the phase transition occurs.

We find that the weak-embedding {\LL} expansion exhibits a behavior that is similar to that of the strong-embedding {\LL} expansion above $T_c$ (see open symbols for $T=1.5$ in the inset in Fig.~\ref{fig:IS_sq_E}). Below $T_c$, like the bond and site expansions, the results from the weak-embedding {\LL} expansion become increasingly different from the exact ones as the order of the expansion increases (see open symbols for $T=0.1$ in the inset in Fig.~\ref{fig:IS_sq_E}). This behavior of the weak-embedding {\LL} expansion, together with the fact that it is computationally more costly than the strong-embedding {\LL} expansion because it has many more clusters, makes the strong-embedding {\LL} expansion our expansion of choice for the Ising model in the calculations that follow. That said, we should mention at $T>T_c$ and sufficiently away from $T_c$, we do find that the weak-embedding {\LL} expansion is slightly more accurate than the strong-embedding {\LL} expansion. This can be seen at $T=1.5$ in the inset in Fig.~\ref{fig:IS_sq_E} if one compares the errors at the highest orders of both expansions.

\begin{figure}[!t]
\includegraphics[width=.98\columnwidth]{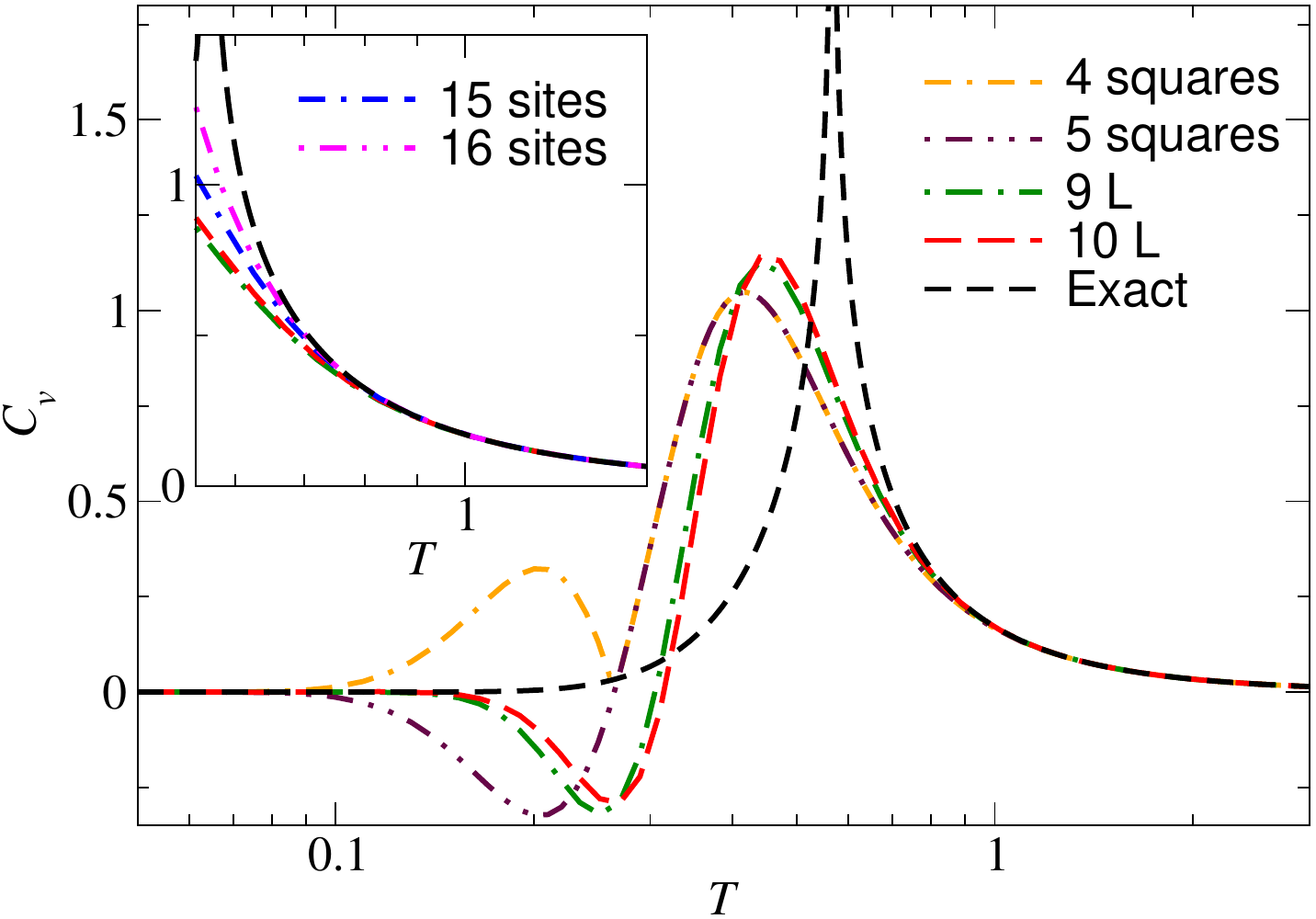}
\caption{Specific heat ($C_v$) vs $T$ for the strong-embedding {\LL} expansion with up to 9 and 10 {\LL}s, the square expansion from Ref.~\cite{rigol_bryant_07a} with up to 4 and 5 squares, and the exact result for the square lattice Ising model. The divergence in the exact result marks $T_c$. (inset) Results for the strong-embedding {\LL} expansion for $T>T_c$, compared to those of the site expansion with up to 15 and 16 sites~\cite{rigol_bryant_07a}, and with the exact result.}\label{fig:IS_sq_Cv}
\end{figure}

In Fig.~\ref{fig:IS_sq_Cv}, we compare the strong-embedding {\LL} expansion and exact results for the specific heat. Like for the energy in Fig.~\ref{fig:IS_sq_E}, the strong-embedding {\LL} expansion for $C_v$ converges at both sides of $T_c$, which is marked by the divergence of the exact result. Figure~\ref{fig:IS_sq_Cv} also shows that the region over which the strong-embedding {\LL} expansion converges improves with increasing the order of the expansion. In Fig.~\ref{fig:IS_sq_Cv}, we also plot results for $C_v$ obtained in Ref.~\cite{rigol_bryant_07a} using the square expansion. They too converge at low and high temperatures, but not as close to $T_c$ as the strong-embedding {\LL} expansion does. In the inset in Fig.~\ref{fig:IS_sq_Cv}, we compare results for the strong-embedding {\LL} expansion and the site expansion with the exact results for $T>T_c$, which is the only regime in which the site expansion converges. They are all similar, though the results for the site expansion follow the exact results to slightly lower temperatures than those of the {\LL} expansion. Overall, one can conclude from the results in Fig.~\ref{fig:IS_sq_Cv} that for the Ising model the strong-embedding {\LL} expansion is a better choice than the site expansion because the {\LL} expansion converges below $T_c$, and than the square expansion because the {\LL} expansion converges over a wider range of temperatures.

Like on the square lattice, the Ising model is exactly solvable on the triangular lattice~\cite{Wannier_1950, HOUTAPPEL1950425, Stephenson_1970}. For an antiferromagnetic coupling ($J>0$), our case of interest here, the triangular lattice Ising model does not develop long-range order at any temperature. This is a consequence of geometric frustration and results in a massive degeneracy of the ground state, which exhibits an entropy per site $S\approx 0.323$ and power-law decaying spin correlations~\cite{Wannier_1950}.

\begin{figure}[!t]
\includegraphics[width=.98\columnwidth]{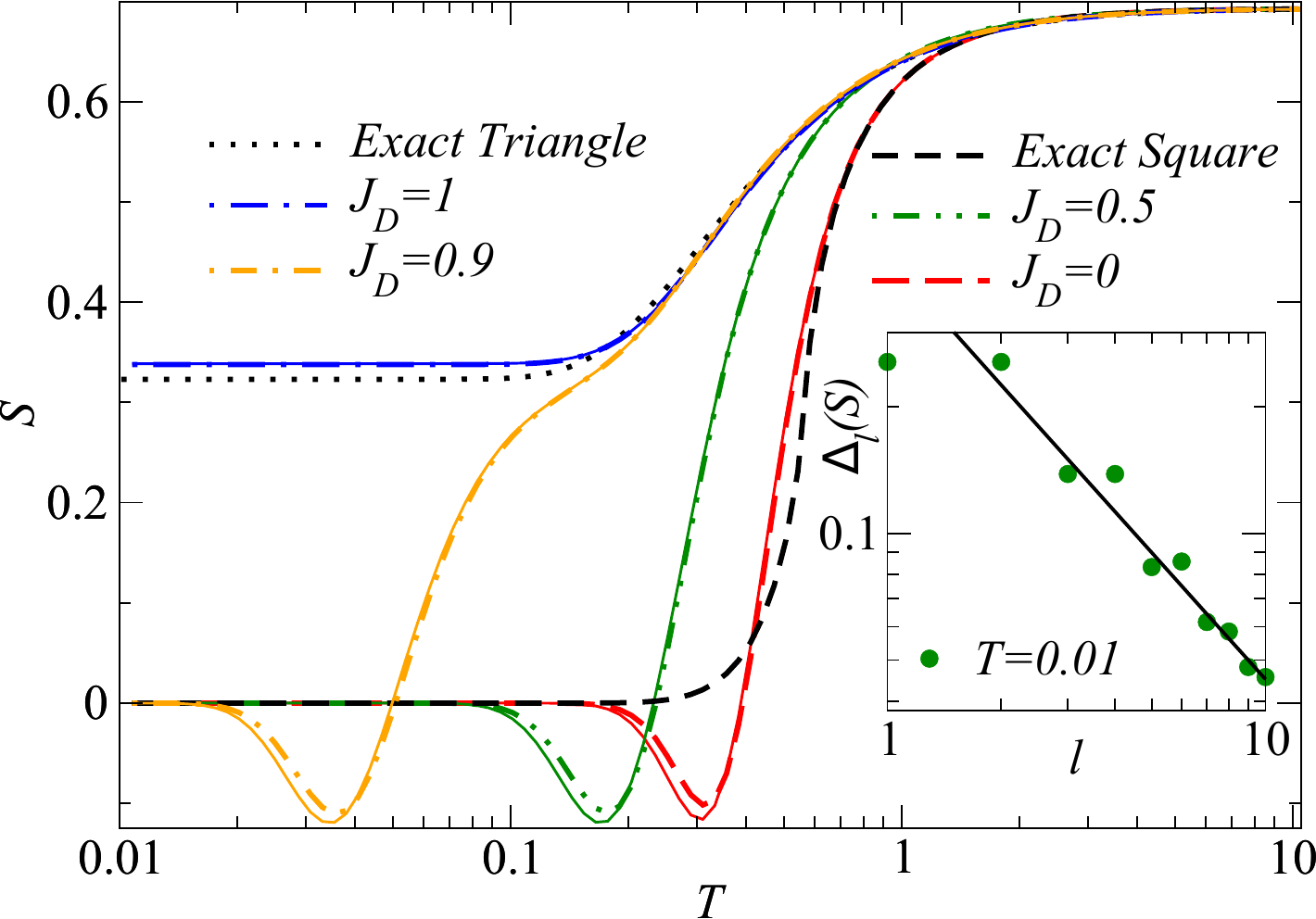}
\caption{Entropy per site ($S$) for the strong-embedding {\LL} expansion as a function of the temperature for different values of the strength of diagonal bonds $J_{D}$. The thin and thick lines show results for up to 9 and 10 {\LL}s, respectively. Exact results are reported for the square, i.e.,~$J_{D}=0$, and triangular, i.e.,~$J_{D}=1$, lattices. (inset) Normalized difference between the entropy obtained for the triangular lattice using the {\LL} expansion and the exact result at $T=0.01$ plotted as a function of the order of the {\LL} expansion. The straight line is a guide to the eye and shows $1/\l$ behavior.}\label{fig:Entropy_IS}
\end{figure}

As mentioned in Sec.~\ref{sec:Ham}, we can study a model that smoothly connects the Ising model on the square and triangular lattice geometries, via adding a diagonal bond of strength $J_D$ to each square in the square lattice and carrying out an {\LL} expansion. So long as $J_D<J$, such a model will still order, because the diagonal bond is weaker and cannot prevent antiferromagnetic order, but it orders below a critical temperature that decreases as $J_D$ approaches $J$. Consequently, the ground state of that model is non-degenerate so long as $J_D<J$. In Fig.~\ref{fig:Entropy_IS}, we show the entropy per site as a function of the temperature for different values of $J_{D}$, as obtained using the strong-embedding {\LL} expansion with up to 9 (thin continuous lines) and 10 (thick lines) {\LL}s. In Fig.~\ref{fig:Entropy_IS}, we also show the exact results for the square and triangular lattice geometries. At the lowest temperatures, the entropy can be seen to vanish for all values of $J_D<J$, and to be nonzero for $J_D=J$. The numerical results for the entropy of the square lattice Ising model agree with the exact results away from $T_c$, as seen before for the energy and the specific heat. Remarkably, because of the lack of long-range order, the numerical results for the triangular lattice Ising model are very close to the exact ones at all temperatures, and exhibit a slightly larger entropy at $T=0$. In the inset in Fig.~\ref{fig:Entropy_IS}, we show that the difference between the low-temperature results for the entropy predicted by the strong-embedding {\LL} expansion and the exact results is consistent with a $1/l$ decrease with increasing the order $l$ of the {\LL} expansion (the straight line shows $1/l$ behavior).

\subsection{Transverse-Field Ising Model}

To investigate whether the observed convergence below $T_c$ for the classical Ising model is extended to quantum models with finite-temperature phase transitions, we study the transverse field Ising model in the square lattice. This model is described by the Hamiltonian in Eq.~(\ref{eq:HIsing}) with $g\neq0$. A sketch of its phase diagram is shown in Fig.~\ref{fig:TFIM_phase}. 

\begin{figure}[!b]
\includegraphics[width=.98\columnwidth]{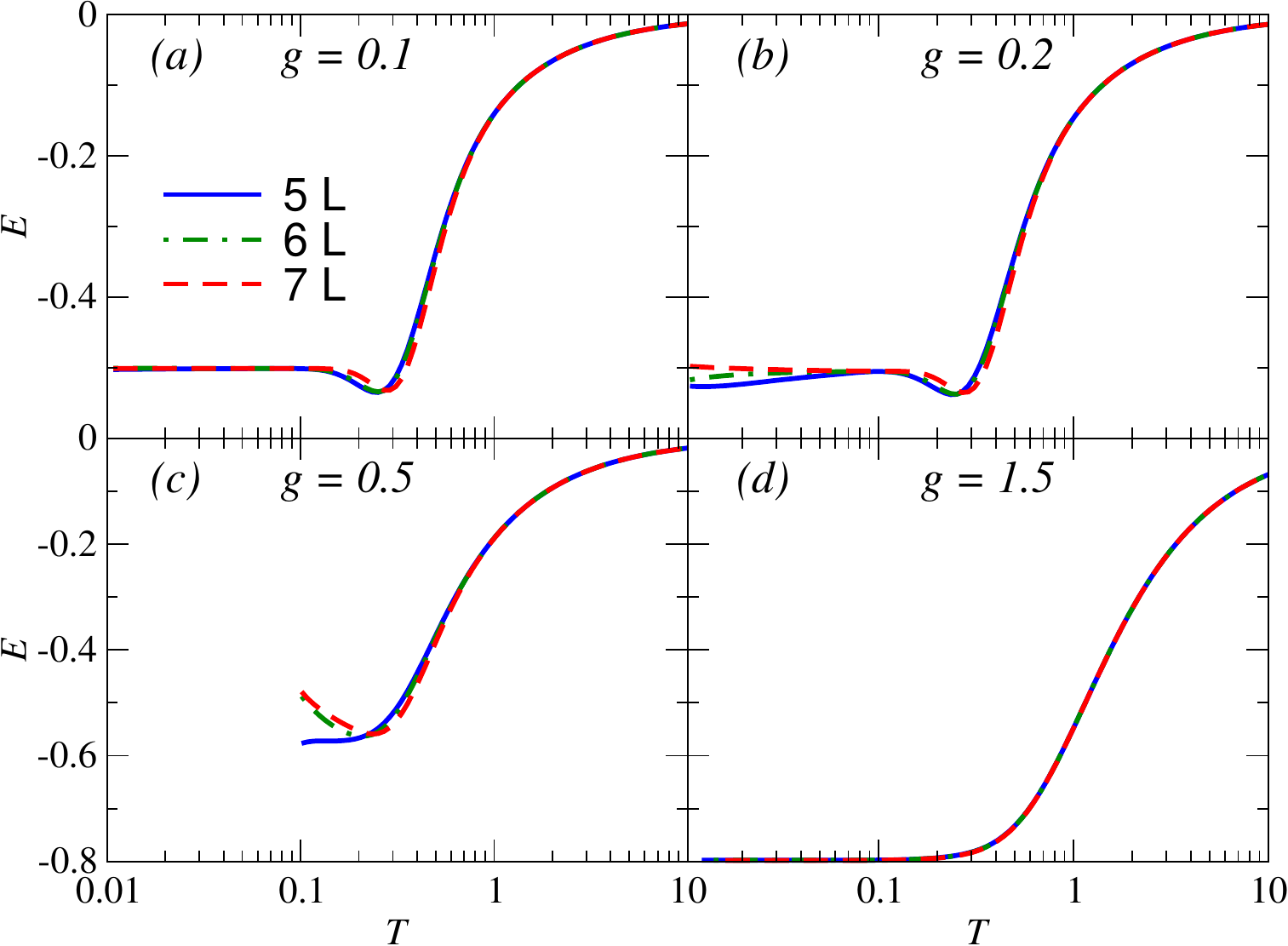}
\caption{Energy $E$ vs $T$ for TFIM on the square lattice with (a) $g = 0.1$, (b) $g = 0.2$, (c) $g = 0.5$ and (d) $g = 1.5$. The highest 3 orders of the {\LL} expansion computed are shown in each panel.} \label{fig:Energy_TFIM_sq}
\end{figure}

In Fig.~\ref{fig:Energy_TFIM_sq}, we show the highest 3 orders of the strong embedding {\LL} expansion for the energy vs temperature for different values of the transverse field strength $g$. We note that the highest order of the {\LL} expansion that we are able to calculate for this model is the lowest of all models considered in this work. It is lower than for the classical Ising model, which is diagonal in the computational basis, because we need to diagonalize the Hamiltonian of the transverse field model in each cluster. It is lower than for the XX and Heisenberg models, which we study in the next sections, because of the absence of $U(1)$ symmetry. Namely, the total $z$-magnetization is not conserved in the transverse field Ising model, so it cannot be used to block diagonalize the Hamiltonian as we do for the XX and Heisenberg models.

The convergence above and below $T_c$ observed in the classical Ising model also occurs after introducing quantum fluctuations through a transverse field of strength $g=0.1$ [see Fig.~\ref{fig:Energy_TFIM_sq}(a)]. Increasing the strength of the transverse field beyond $g=0.1$, see results for $g=0.2$ in Fig.~\ref{fig:Energy_TFIM_sq}(b) and for $g=0.5$ in Fig.~\ref{fig:Energy_TFIM_sq}(c), we find that higher orders of the {$\LL$} expansion are needed for the bare sums to converge in the scale of the plots. For $g=0.5$, we find that the convergence is lost at the lowest temperatures for the orders of the {$\LL$} expansion that we are able to calculate. Still, the results for the energy converge well below $T=1$, which is where one expects high-temperature expansions to diverge. For $g \gtrsim g_c$, since the system is disordered, the NLCE results converge at all temperatures [see Fig.~\ref{fig:Energy_TFIM_sq}(d) for $g=1.5$]. For models with moderate to strong fields, which are in many instances of relevance to understanding experimental results, NLCEs can provide very accurate results independently of whether there is or not frustration in paramagnetic phases.

\begin{figure}[!t]
\includegraphics[width=.98\columnwidth]{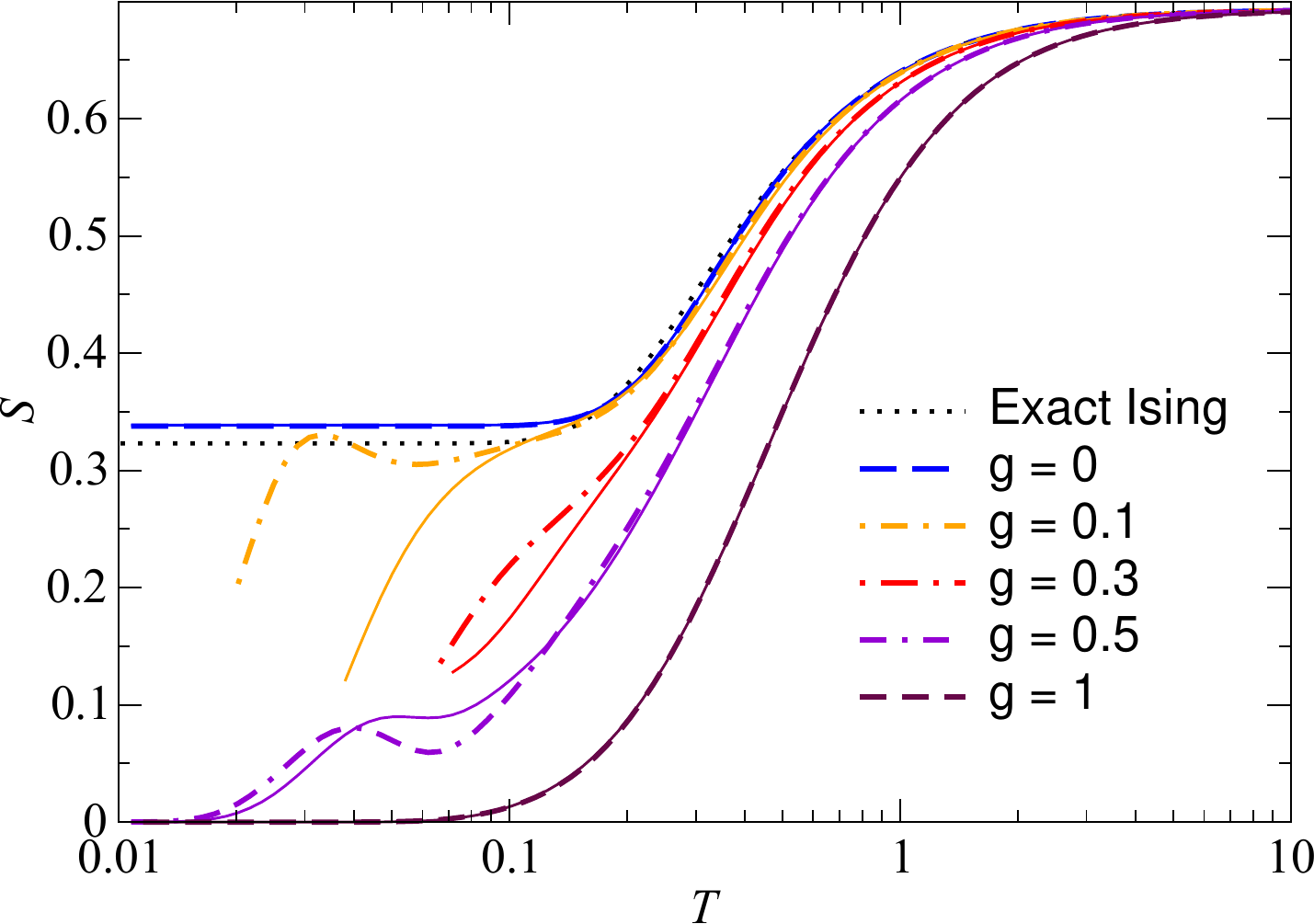}
\caption{Entropy $S$ vs $T$ for the transverse field Ising model in the triangular lattice. We show results obtained with the strong embedding {$\LL$} expansion with up to 9 (thin lines) and 10 (thick lines) {\LL}s for $g=0$, and up to 6 (thin lines) and 7 (thick lines) {\LL}s for $g=0.1$, 0.3, 0.5, and 1. Exact results are reported for the classical Ising model ($g=0$).} \label{fig:Entropy_TFIM_tr}
\end{figure}

The addition of a transverse field to the Ising model in the triangular lattice lifts the massive degeneracy of the ground state present in the classical case. The quantum fluctuations introduced by the field pick a symmetry-broken (bond-ordered) ground state through an ``order-from-criticality'' mechanism~\cite{Moessner_2000, Moessner_2001}. This phase persists up to a critical field $g_c \approx 0.82$, after which the ground state is paramagnetic~\cite{Moessner_2003}. In the bond-ordered phase, increasing the temperature results in two BKT transitions that bound a BKT phase~\cite{Moessner_2001, Moessner_2003}. In Fig.~\ref{fig:Entropy_TFIM_tr}, we show results for the entropy of the transverse field Ising model in the triangular lattice as one increases the strength of the field $g$, and the model departs from the classical $g=0$ limit. At the weakest ($g=0.1$) field shown, the entropy follows the classical result at high and intermediate temperatures and then departs at low temperatures showing indications that it vanishes at the lowest temperatures. As expected, increasing the strength of the field results in a departure from the classical result occurring at higher temperatures. For $g\gtrsim g_c$, we again find that the NLCE results converge at all temperatures. Like in Fig.~\ref{fig:Energy_TFIM_sq}, the results in Fig.~\ref{fig:Entropy_TFIM_tr} show that NLCEs are an excellent choice when it comes to obtaining accurate results for thermodynamic properties of quantum lattice models in the presence of magnetic fields at low temperatures.

\section{XX Model}\label{sec:xx}

\begin{figure}[!b]
\includegraphics[width=.48\textwidth]{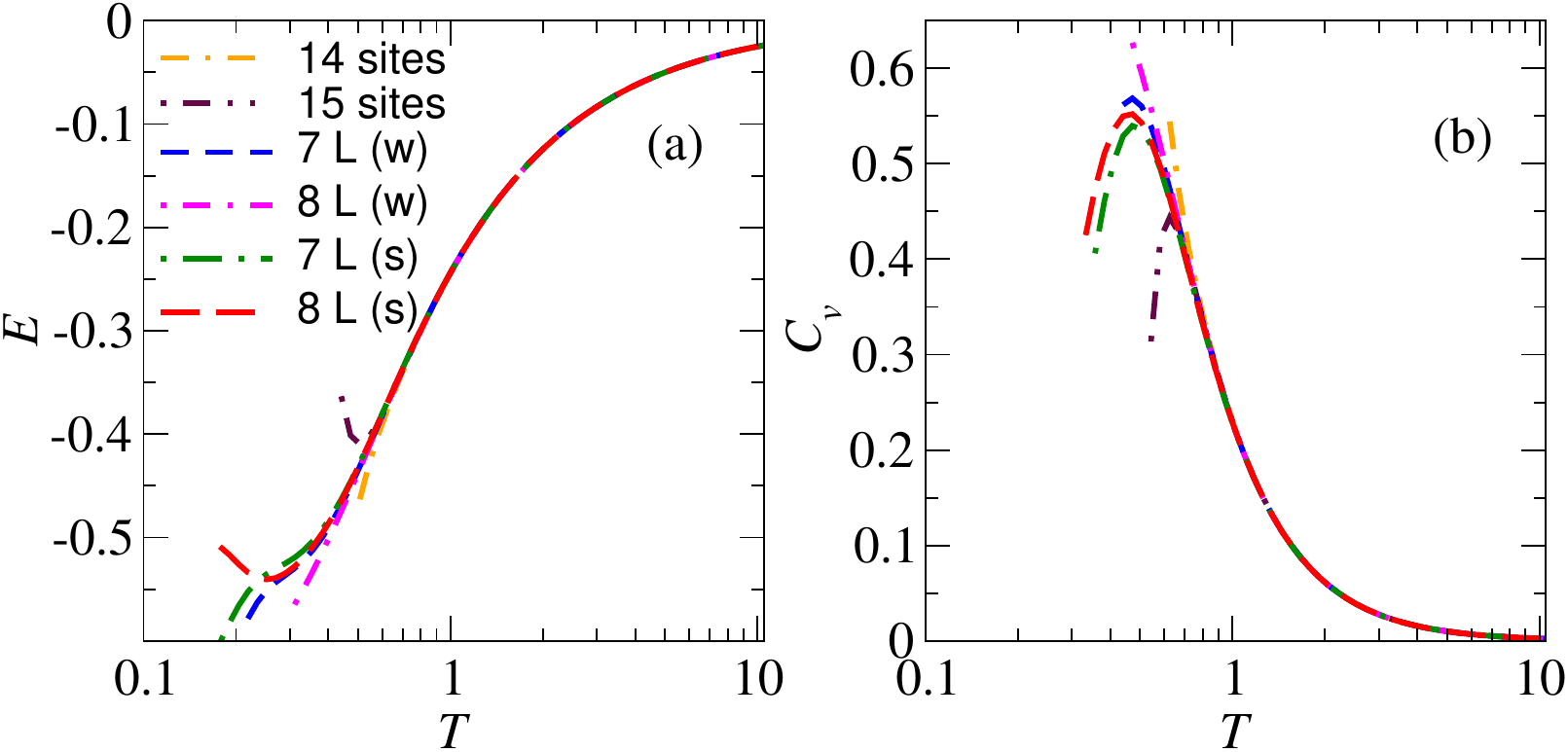}
\caption{(a) Energy and (b) specific heat vs $T$ for the XX model in the square lattice. We show results for the weak (w) and strong (s) embedding {\LL} expansions with up to 7 and 8 {\LL}s, as well as for the site expansion with up to 14 and 15 sites.} \label{fig:XY_E_Cv_sq}
\end{figure}

As mentioned in Sec.~\ref{sec:Ham}, in the square lattice the XX model does not exhibit long-range order at nonzero temperatures, and develops quasi-long-range order at temperatures below the BKT transition temperature $T_{KT}\approx0.343$. In Fig.~\ref{fig:XY_E_Cv_sq}(a), we show NLCE results for the energy plotted as a function of the temperature. We compare results obtained using the site expansion with up to 14 and 15 sites, with results for the weak and strong embedding {\LL} expansions with up to 7 and 8 {\LL}s. The results for the two orders of the site expansion agree with each other at temperatures $T\gtrsim 0.6$, while those for the {\LL} expansions agree with each other at $T\gtrsim 0.4$, i.e., the bare sums of the {\LL} expansion converge at lower temperatures ($T\gtrsim T_{KT}$) than those of the site expansion. We also find that the temperatures at which the last two orders of the strong embedding {\LL} expansion start to agree with each other are slightly lower than those for the weak embedding {\LL} expansion.

In Fig.~\ref{fig:XY_E_Cv_sq}(b), we show NLCE results for the specific heat for the same orders of the expansions as those for the energy in Fig.~\ref{fig:XY_E_Cv_sq}(a). The NLCE results for the specific heat parallel those for the energy, with the different orders of the {\LL} expansions agreeing with each other at lower temperatures than those for the site expansion, and with the strong embedding {\LL} expansion converging at a slightly lower temperature than the weak embedding {\LL} expansion. Given the lower computational cost of the strong embedding {\LL} expansion, the results in Fig.~\ref{fig:XY_E_Cv_sq} make this expansion the expansion of choice to study this model, and the Heisenberg model in the next section. 

Two general comments to be made at this point are that: (i) As expected~\cite{rigol_bryant_07a}, in Fig.~\ref{fig:XY_E_Cv_sq} one can see that the NLCE results for the energy converge to lower temperatures than those for the specific heat and, in general, than for more complicated observables. (ii) Because of the presence of the BKT transition, and similarly for the Ising model because of the presence of the order-to-disorder phase transition, resummation techniques do not allow one to significantly extend the region of convergence of the NLCE results beyond those of the bare sums. Hence, we postpone the use of resummation techniques to the next section on the Heisenberg model.

\begin{figure}[!t]
\includegraphics[width=.98\columnwidth]{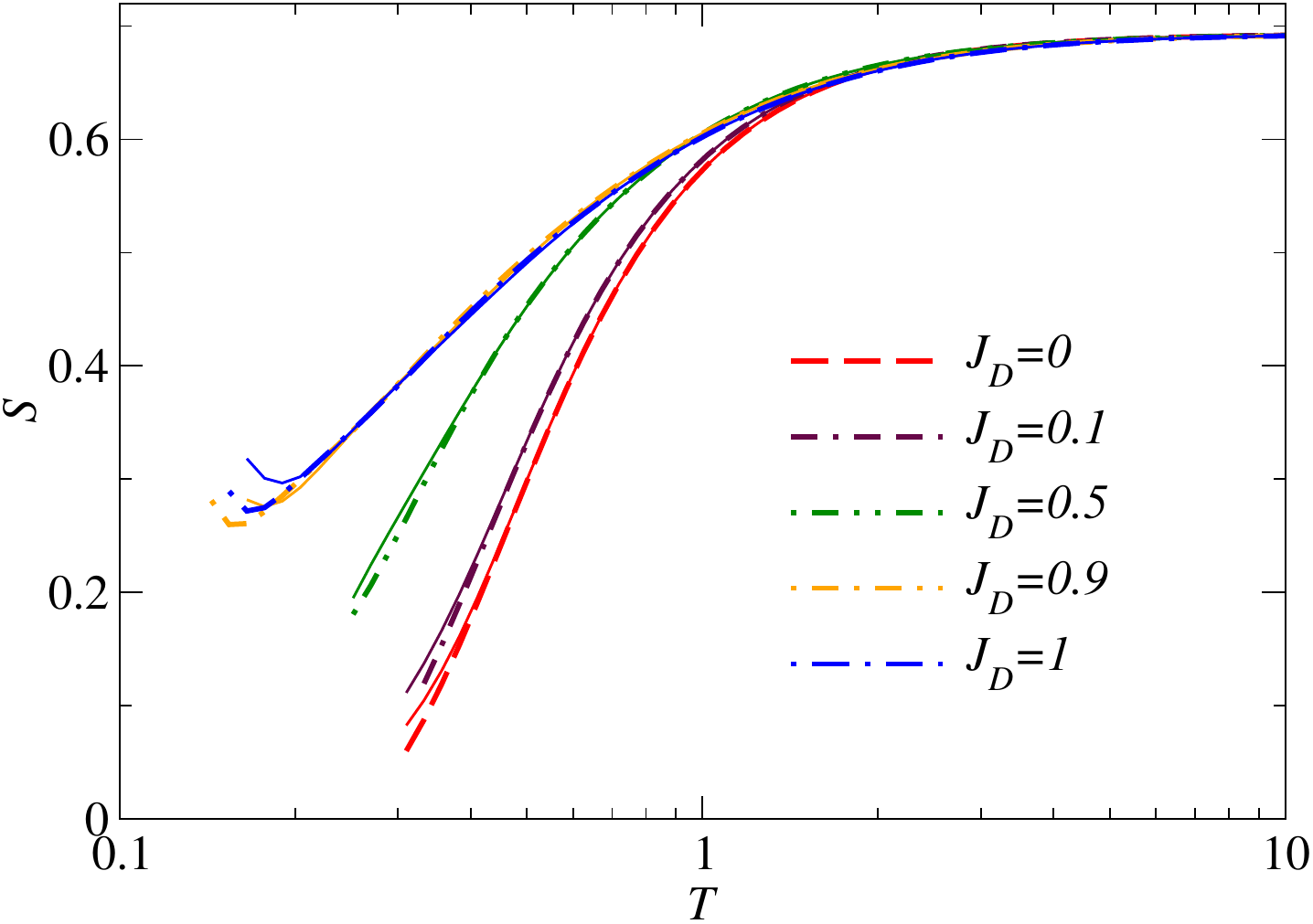}
\caption{Strong embedding {\LL} expansion results for the entropy of the XX model as a function of the temperature obtained from different values of $J_D$, namely, as one transitions between the square, i.e.,~$J_D=0$, and the triangular, i.e., $J_D=1$, lattices. The thin and thick lines show results for up to 7 and 8 {\LL}s, respectively.} \label{fig:XY_sq_tr}
\end{figure}

We close this section on the XX model exploring the transition between the square and the triangular lattice geometries. In the triangular lattice, the antiferromagnetic XX model is frustrated. The ground state exhibits a two-fold degeneracy and, at finite temperature, the model exhibits two transitions, a BKT transition and an Ising transition~\cite{Shiba_XY, Landau_XY_tr}. In Fig.~\ref{fig:XY_sq_tr}, we plot the entropy as a function of the temperature for different strengths $J_D$ of the diagonal bonds. As for the Ising model, in Fig.~\ref{fig:Entropy_IS}, in Fig.~\ref{fig:XY_sq_tr} we can see a hallmark of frustration. At low temperatures, the entropy in the triangular lattice is higher than in the square lattice. Namely, as a result of frustration, more states are accessible in the former geometry at low temperatures, and Fig.~\ref{fig:XY_sq_tr} shows how the entropy increases for $T<1$ as $J_D$ departs from 0. One can also see in Fig.~\ref{fig:XY_sq_tr} that, since at any given temperature frustration shortens the range of the correlations, the NLCE results for the entropy converge at lower temperatures as $J_D$ increases from 0 to 1.

\section{Heisenberg model}\label{sec:heisenberg}

The ground state of the antiferromagnetic Heisenberg model on the square lattice is known to exhibit long-range antiferromagnetic correlations~\cite{AFHM_GS_Singh, AFHM_GS_Troyer}. This model does not exhibit a finite temperature transition, but antiferromagnetic correlations are known to grow rapidly as $T$ decreases below $T=1$. Because of the latter behavior, the bare sums are not expected to converge up to temperatures as low as for the XX model. However, due to the absence of a phase transition, resummation techniques can provide insights into the behavior of thermodynamic quantities at significantly lower temperatures than those at which the bare sums converge.

\begin{figure}[!b]
\includegraphics[width=.98\columnwidth]{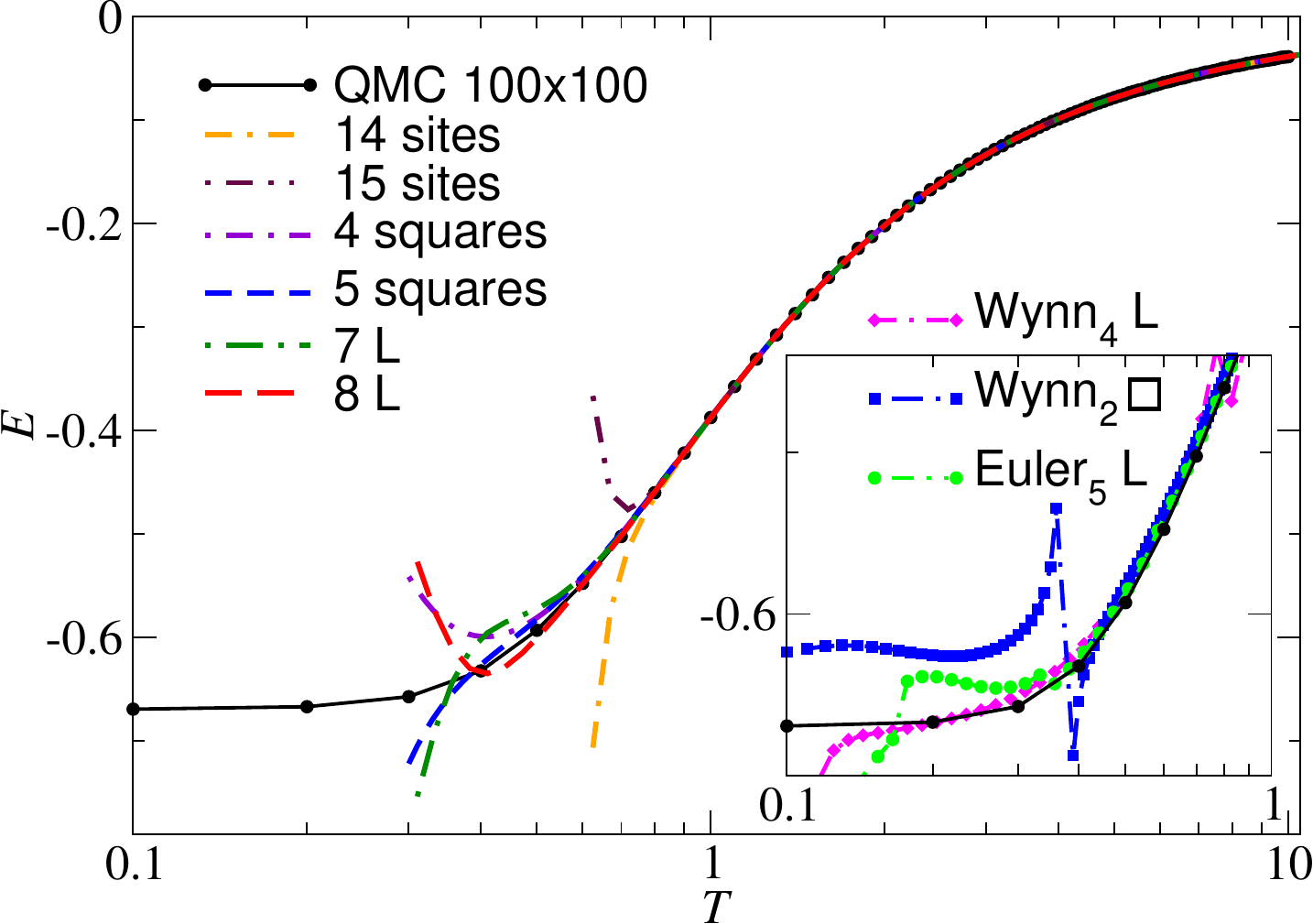}
\caption{Energy $E$ vs temperature $T$ for the antiferromagnetic Heisenberg model on the square lattice. We compare the results obtained from the bare sums for the site, the square, and the strong embedding {\LL} expansion (as per the legend), with QMC results for a lattice with 100 $\times$ 100 sites. (inset) Comparison between the highest order (4) of the Wynn resummation for the strong embedding {\LL} expansion (Wynn$_4$~{\LL}), the highest order (2) of the Wynn resummation for the square expansion (Wynn$_2~\square$), Euler$_5$ for the {\LL} expansion [see Eq.~\eqref{eq:euler}], and the QMC results.} \label{fig:AF_E_bare}
\end{figure}

In Fig.~\ref{fig:AF_E_bare}, we show results obtained for the energy of the Heisenberg model as a function of the temperature using the bare sums of the site, square, and the strong embedding {\LL} expansions. The NLCE results are compared to QMC results obtained using the SSE technique~\cite{R_Yu, rigol_bryant_07a}, which can be thought of as being the exact results at those temperatures. Despite having substantially different number of clusters and orders included in the expansion, the results for the square and the strong embedding {\LL} expansions are qualitatively similar. They converge to lower temperatures than those of the site expansion. When converged, i.e., when in agreement, all the NLCE results agree with the QMC ones. At the scale of the plot, the two orders of the strong embedding {\LL} expansion overlap with each other and with the QMC results at temperatures $T \gtrsim 0.6$. The inset in Fig.~\ref{fig:AF_E_bare} shows that order 4 of Wynn's resummation for the the \LL~expansion produces energies that are close to the QMC results, and the agreement improves with increasing the order of the resummation (other orders not shown for clarity), down to temperatures $T\sim 0.15$. This is similar to what was reported in Ref.~\cite{rigol_bryant_07a} when Wynn's resummation was used for the site expansion. In contrast, the results of order 2 of Wynn's resummation (the highest order accessible) for the square expansion, which are also shown in the inset, make apparent that there are not enough orders of the square expansion for Wynn's resummation to produce significant improvements for this model. The inset in Fig.~\ref{fig:AF_E_bare} further shows that the Euler transformation for the \LL~expansion gives results that are close to the exact ones up to $T\sim 0.35$.

\begin{figure}[!t]
\includegraphics[width=.98\columnwidth]{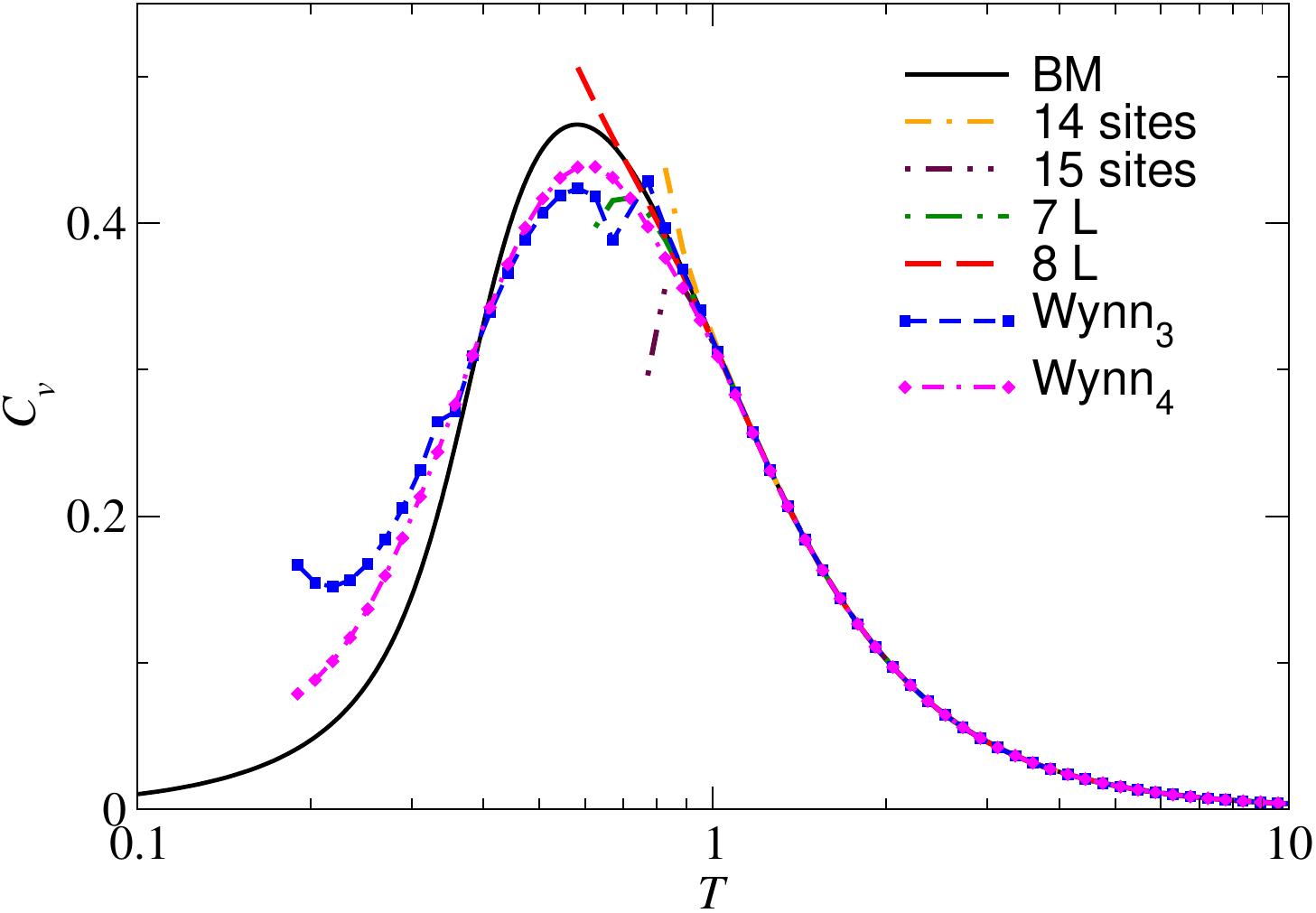}
\caption{Specific heat $C_v$ vs temperature $T$ for the antiferromagnetic Heisenberg model on the square lattice. We compare the results obtained from the bare sums for the site and the strong embedding {\LL} expansion (as per the legend), as well as the highest two orders of the Wynn resummation for the strong embedding {\LL} expansion, with results reported by Bernu and Misguich (BM) in Ref.~\cite{bernu_01}.} \label{fig:AF_CV}
\end{figure} 

Hence, the results in the main panel in Fig.~\ref{fig:AF_E_bare} show that the bare sums of the strong embedding {\LL} expansion share the advantages of the square expansion over the site expansion, namely, a convergence of the bare sums to lower temperatures. The inset in Fig.~\ref{fig:AF_E_bare}, on the other hand, shows that the strong embedding {\LL} expansion shares the advantages of the site expansion over the square expansion when it comes to resummation techniques, namely, resummation techniques allow one to obtain meaningful results at signifficantly lower temperatures than those accessible via the bare sums.

In Fig.~\ref{fig:AF_CV}, we show results for the bare sums of the site and strong embedding {\LL} expansion for the specific heat, and compare them to results reported by Bernu and Misguich (BM) in Ref.~\cite{bernu_01}. For $C_v$, the bare sums of the {\LL} expansion converge only to slightly lower temperatures than the site expansion. The difference is not as marked as for the energy in Fig.~\ref{fig:AF_E_bare}. When in agreement with each other, the NLCE results also agree with those from Ref.~\cite{bernu_01}, but the NLCE results show no hints of the nearby maximum seen in BM results. The Wynn resummation results also reported in Fig.~\ref{fig:AF_CV}, on the other hand, exhibit a maximum at approximately the same temperature as the results in Ref.~\cite{bernu_01}. One can see that the maximum in the Wynn resummation is higher for the highest Wynn order so it is possible that if more orders of the {\LL} expansion were calculated the results from higher order Wynn resummations will agree with the result in Ref.~\cite{bernu_01}. We should add that the Wynn resummation results for the site expansion, reported in Ref.~\cite{rigol_bryant_07a}, behave similarly as those of the {\LL} expansion in Fig.~\ref{fig:AF_CV}.

\begin{figure}[!t]
\includegraphics[width=.98\columnwidth]{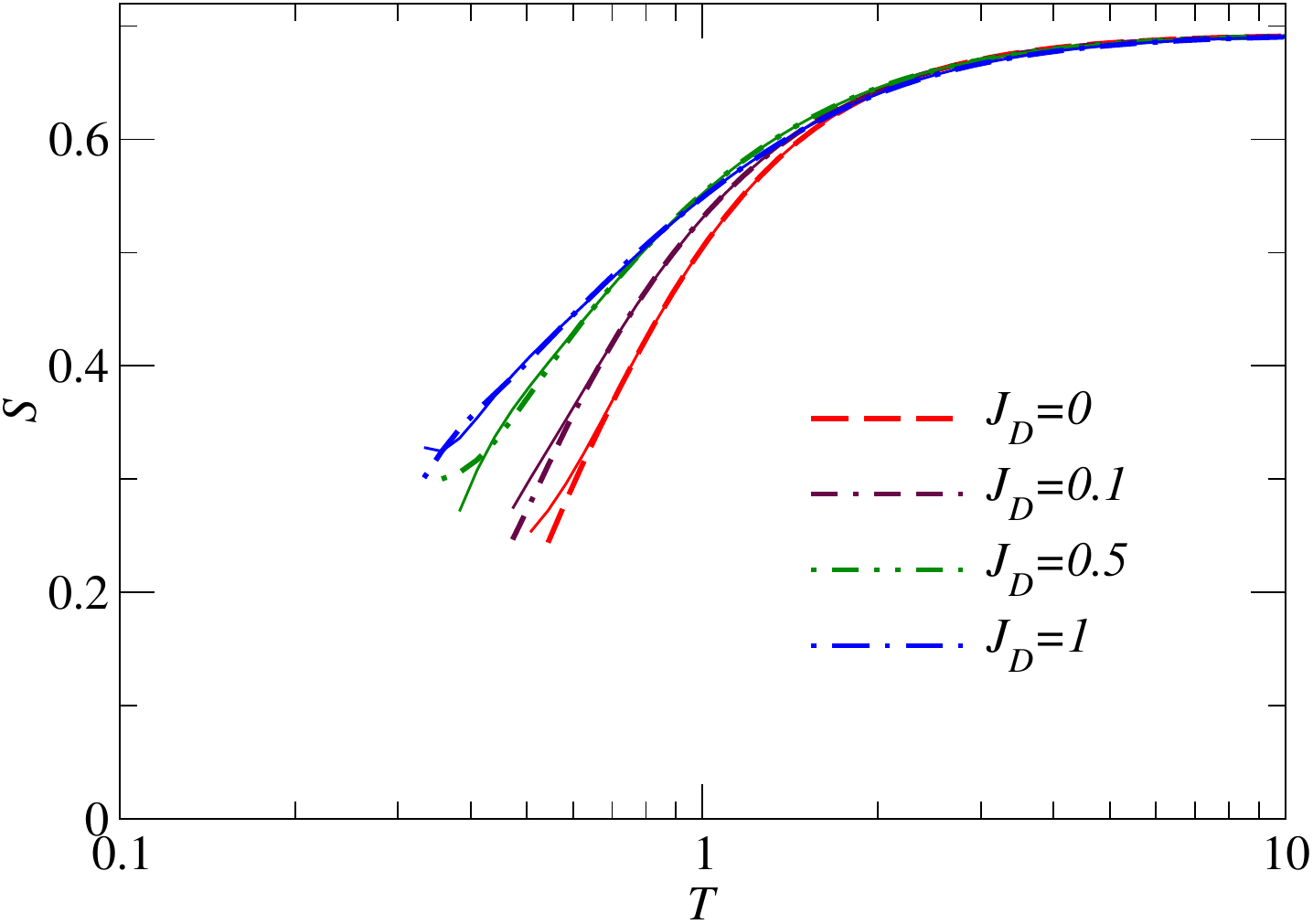}
\caption{Strong embedding {\LL} expansion results for the entropy of the antiferromagnetic Heisenberg model as a function of the temperature obtained from different values of $J_D$, namely, as one transitions between the square, i.e.,~$J_D=0$, and triangular, i.e.,~$J_D=1$, lattice geometries. The thin and thick lines show results for up to 7 and 8 {\LL}s, respectively.} \label{fig:AF_sq_tr}
\end{figure} 

The antiferromagnetic Heisenberg model in the triangular lattice is frustrated and exhibits long-range order at zero temperature~\cite{Bernu_AFHM_tr_GS}. Because of the effect of frustration, the spin-spin correlations at any given temperature are shorter than in the square lattice~\cite{Elstner_AFHM_tr_GS}. In Fig.~\ref{fig:AF_sq_tr}, we plot the entropy as a function of the temperature as the Heisenberg model transitions between the square and the triangular lattice geometries. All the results were obtained using the strong embedding {\LL} expansion. As expected due to frustration, and as we have seen for the Ising and XX models, the entropy at low temperatures increases as $J_{D}$ increases. In parallel, we find that the bare sums converge to lower temperatures also as $J_{D}$ increases. The convergence of the bare sums for the Heisenberg model in Fig.~\ref{fig:AF_sq_tr} is worse than that of the XX model in Fig.~\ref{fig:XY_sq_tr}, i.e., the Heisenberg model is more challenging to study using NLCEs.

\begin{figure}[!t]
\includegraphics[width=.98\columnwidth]{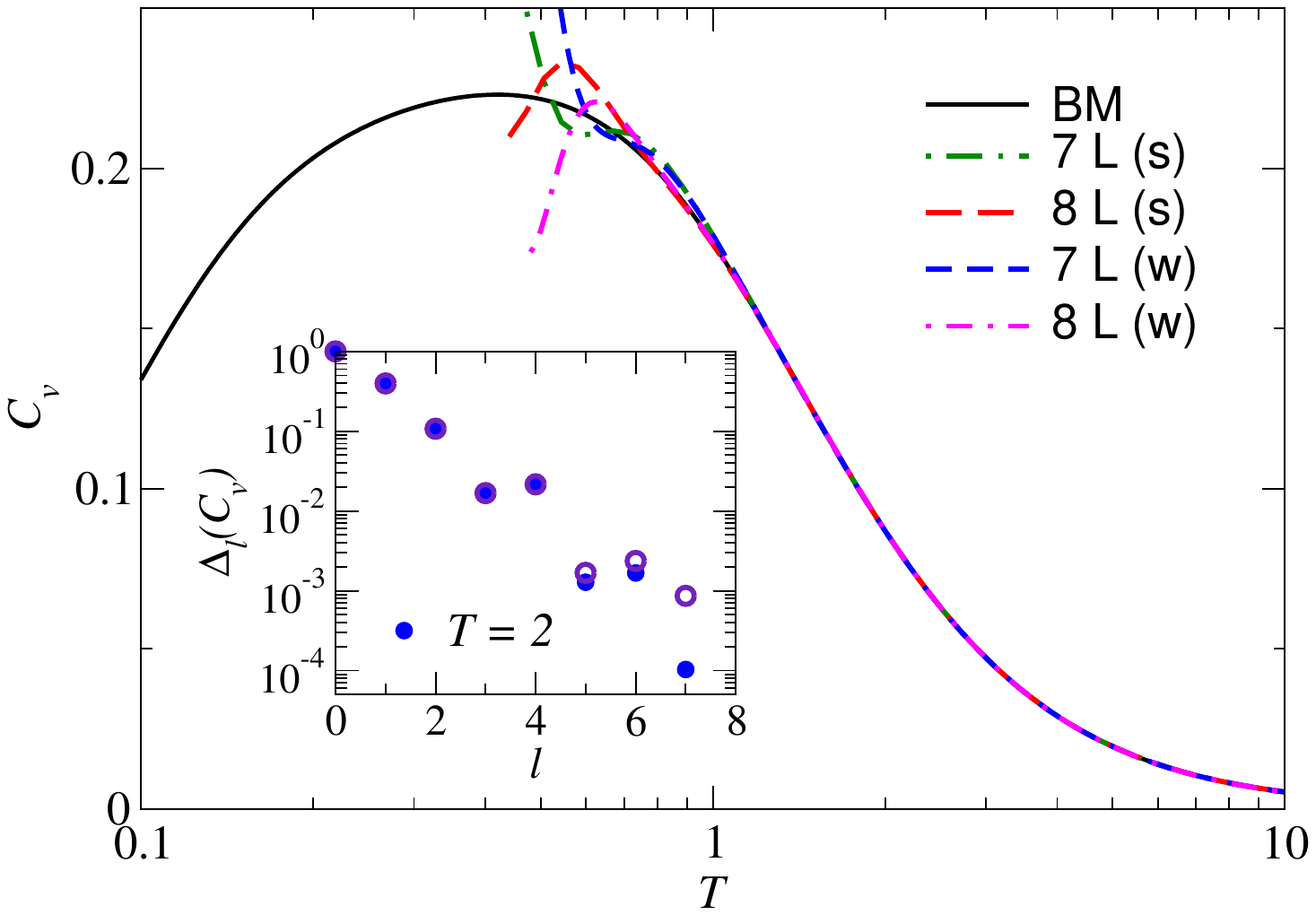}
\caption{Specific heat $C_v$ vs temperature $T$ for the antiferromagnetic Heisenberg model on the triangular lattice. We compare the results obtained from the bare sums for the weak (w) and strong (s) embedding {\LL} expansions with results reported by Bernu and Misguich (BM) in Ref.~\cite{bernu_01}. (inset) Normalized difference between the specific heat obtained using the strong-embedding (filled symbols) and weak-embedding (empty symbols) {\LL} expansions and the highest order (8) computed for each expansion vs the order of the expansion.} \label{fig:AF_CV_tr}
\end{figure}

We conclude this section showing results for the specific heat of the triangular lattice Heisenberg model (see Fig.~\ref{fig:AF_CV_tr}) obtained using the weak and the strong embedding {\LL} expansions. For this specific lattice geometry, one can refer to those expansions as the weak and strong embedding triangle expansions, of which the weak embedding version was introduced in Ref.~\cite{rigol_bryant_07a}. Figure~\ref{fig:AF_CV_tr} shows that both expansions give nearly identical results, despite the fact that the strong embedding expansion includes fewer clusters. When both expansions agree with each other, they also agree with results from BM reported in Ref.~\cite{bernu_01}. In the inset in Fig.~\ref{fig:AF_CV_tr}, we show that both expansions exhibit similar convergence properties for $T\gtrsim 1$. Hence, we plan to use the strong embedding version in future studies of the triangular lattice.

\section{Summary}\label{sec:summary}

We introduced a NLCE for square-lattice models whose building block is an {\LL}-shape cluster. We showed that this expansion shares the advantages of expansions based on larger building blocks, such as corner-sharing squares, over expansions based on smaller building blocks, such as sites and bonds. Namely, the bare sums of the {\LL} expansion converge to lower temperatures than the site- and bond-based expansions, and exhibit a similar or better convergence than that of the square expansion. We also showed that the {$\LL$} expansion can be carried out to sufficiently high orders to enable the use of resummation techniques to signifficantly extend the convergence beyond that of the bare sums. In this regard, the {$\LL$} expansion shares an advantage of the smaller building blocks bond and site expansions over the square expansion. In addition, we showed that in disordered phases in the presence of magnetic fields (and this holds no matter whether or not there is frustration), the {\LL} expansion can provide accurate results at all temperatures.

We introduced weak-embedding and strong-embedding versions of the {$\LL$} expansion, and showed that they exhibit similar convergence properties at intermediate and high temperatures, while in ordered phases at very low temperatures, only the strong embedding version converges for the models considered here. Given the lower computational cost of the strong embedding {$\LL$} expansion, which has fewer topologically distinct clusters at any given order of the expansion, and the previously mentioned convergence properties make the strong embedding {$\LL$} expansion the more appealing one of the two for future studies. Finally, we showed that the expansion based on the {\LL}-shape cluster can be naturally used to study properties of lattice models that smoothly connect the square and triangular lattice geometries.

\acknowledgments
This work was supported by the National Science Foundation (NSF) Grant No.~PHY-2012145.

\bibliographystyle{biblev1}
\bibliography{Reference}

\end{document}